\documentclass[a4paper]{article}
\usepackage{graphicx,amsmath,amssymb}
\newcommand{\haak}[1]{\!\left(#1\right)}
\newcommand{\rhaak}[1]{\!\left [#1\right]}
\newcommand{\haaksp}[1]{\left(#1\right)}
\newcommand{\rhaaksp}[1]{\left [#1\right]}
\newcommand{\lhaak}[1]{\left | #1\right |}
\newcommand{\ahaak}[1]{\!\left\{#1\right\}}
\newcommand{\gem}[1]{\left\langle #1\right\rangle}

\newcommand{\rhaakl}[1]{\left[#1\right.}
\newcommand{\rhaakr}[1]{\left.#1\right]}

\newcommand{\lhaakr}[1]{\left.#1\right |}

\newcommand{\floor}[1]{\left\lfloor #1\right\rfloor}
\newcommand{\half}{\frac{1}{2}}

\newcommand{\bfm}[1]{\mathbf{#1}}
\renewcommand{\imath}{\text{i}}
\renewcommand{\dfrac}[2]{#1/#2}
\newtheorem{conjecture}{Conjecture}
\newtheorem{assumption}{Assumption}

\begin{document}
\title{Exact asymptotics of the characteristic polynomial of the symmetric Pascal matrix} 
\author{Saibal Mitra\\
Instituut voor Theoretische Fysica\\
Universiteit van Amsterdam\\
1018 XE Amsterdam\\
The Netherlands}
\date{\today}
\maketitle
\begin{abstract}
We have obtained the exact asymptotics of the determinant \newline $\det_{1\leq r,s\leq L}\rhaak{\binom{r+s-2}{r-1}+\exp\haak{i\theta}\delta_{r,s}}$. Inverse symbolic computing methods were used to obtain exact analytical expressions for all terms up to relative order $L^{-14}$ to the leading term. This determinant is known to give weighted enumerations of cyclically symmetric plane partitions, weighted enumerations of certain families of vicious walkers and it has been conjectured to be proportional to the one point function of the O$\haaksp{1}$ loop model on a cylinder of circumference $L$. We apply our result to the loop model and give exact expressions for the asymptotics of the average of the number of loops surrounding a point and the fluctuation in this number. For the related bond percolation model at the critical point, we give exact expressions for the asymptotics of the probability that a point is on a cluster that wraps around a cylinder of even circumference and the probability that a point is on a cluster spanning a cylinder of odd circumference.

\end{abstract}
\section{Introduction}
The binomial determinant
\begin{equation}\label{dlth}
\det_{1\leq r,s\leq L}\rhaak{\binom{r+s-2}{r-1}+\exp\haak{\imath\theta}\delta_{r,s}}
\end{equation}
gives a weighted enumeration of certain types of nonintersecting lattice paths and cyclically symmetrical plane partitions \cite{lozenge}.
Consider a family of nonintersecting lattice paths on an $L\times L$ square lattice subjected to the constraints:
\begin{enumerate}
\item Paths are allowed to move one step to the right or downward.
\item Paths are required to start from a point $\haak{0,k}$ with $0\leq k\leq L-1$.
\item A path starting at $\haak{0,k}$ is required to end at the point $\haak{k,0}$.
\end{enumerate}
See Fig.\ \ref{fig:path} for an example.
\setlength{\unitlength}{0.8\textwidth}
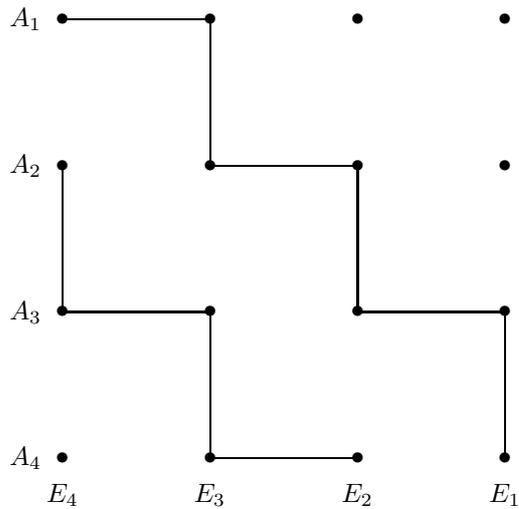
\begin{figure}
\begin{center}
\begin{picture}(0.6,0.6)(-0.08,-0.08)
\put(0,0){\makebox(0,0){$\bullet$}}
\put(0.2,0){\makebox(0,0){$\bullet$}}
\put(0.4,0){\makebox(0,0){$\bullet$}}
\put(0.6,0){\makebox(0,0){$\bullet$}}

\put(0,0.2){\makebox(0,0){$\bullet$}}
\put(0.2,0.2){\makebox(0,0){$\bullet$}}
\put(0.4,0.2){\makebox(0,0){$\bullet$}}
\put(0.6,0.2){\makebox(0,0){$\bullet$}}

\put(0,0.4){\makebox(0,0){$\bullet$}}
\put(0.2,0.4){\makebox(0,0){$\bullet$}}
\put(0.4,0.4){\makebox(0,0){$\bullet$}}
\put(0.6,0.4){\makebox(0,0){$\bullet$}}

\put(0,0.6){\makebox(0,0){$\bullet$}}
\put(0.2,0.6){\makebox(0,0){$\bullet$}}
\put(0.4,0.6){\makebox(0,0){$\bullet$}}
\put(0.6,0.6){\makebox(0,0){$\bullet$}}

\put(-0.05,0){\makebox(0,0){$A_{4}$}}
\put(-0.05,0.2){\makebox(0,0){$A_{3}$}}
\put(-0.05,0.4){\makebox(0,0){$A_{2}$}}
\put(-0.05,0.6){\makebox(0,0){$A_{1}$}}
\put(0,-0.05){\makebox(0,0){$E_{4}$}}
\put(0.2,-0.05){\makebox(0,0){$E_{3}$}}
\put(0.4,-0.05){\makebox(0,0){$E_{2}$}}
\put(0.6,-0.05){\makebox(0,0){$E_{1}$}}
\put(0,0.6){\line(1,0){0.2}}
\put(0.2,0.6){\line(0,-1){0.2}}
\put(0.2,0.4){\line(1,0){0.2}}
\put(0.4,0.4){\line(0,-1){0.2}}
\put(0.4,0.2){\line(1,0){0.2}}
\put(0.6,0.2){\line(0,-1){0.2}}
\put(0,0.4){\line(0,-1){0.2}}
\put(0,0.2){\line(1,0){0.2}}
\put(0.2,0.2){\line(0,-1){0.2}}
\put(0.2,0){\line(1,0){0.2}}
\end{picture}
\caption{A typical configuration of nonintersecting lattice paths enumerated by the determinant in \eqref{dlth} for $L=4$. Paths starting at $A_{i}$ have to end at $E_{i}$.}\label{fig:path}
\end{center}
\end{figure}
The determinant \eqref{dlth} gives a weighted enumeration of all possible lattice paths satisfying the above constraints, where a configuration containing $s$ paths is given a weight of $\exp\rhaak{\imath\theta \haak{L-s}}$.

Families of lattice paths of this type are in bijection with cyclically symmetrical plane partitions \cite{bij,lozenge}. A plane partition of an integer $N$ is an array of nonnegative integers $n_{j,k}$, such that $n_{j+1,k}\leq n_{j,k}$, $n_{j,k+1}\leq n_{j,k}$ and $N=\sum_{j=1}^{\infty}\sum_{k=1}^{\infty}n_{j,k}$. Plane partitions can be represented by a pile of unit cubes by introducing $x$,$y$,$z$ coordinates in $\mathbb{Z}^{3}$ and placing at position $(j,k,0)$ a stack of $n_{j,k}$ unit cubes. A cyclically symmetric plane partition is a plane partition whose representation as a pile of cubes is symmetric under a cyclic permutation of the $x$,$y$,$z$ coordinates. The bijection maps lattice paths of the above type to cyclically symmetric plane partitions that fit in an $L\times L\times L$ box. Families of lattice paths containing $n$ paths are mapped to cyclically symmetric plane partitions with $n$ cubes on the main diagonal.

\begin{figure}[t]
\begin{center}
\includegraphics[width=0.5\textwidth]{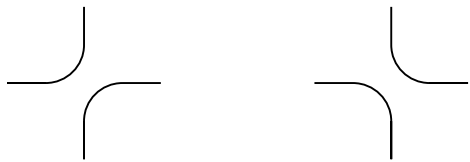}
\caption{The two vertices of the dense O(1) loop model}\label{fig:vrt}
\end{center}
\end{figure}

\begin{figure}[ht]
\begin{center}
\includegraphics[width=0.5\textwidth]{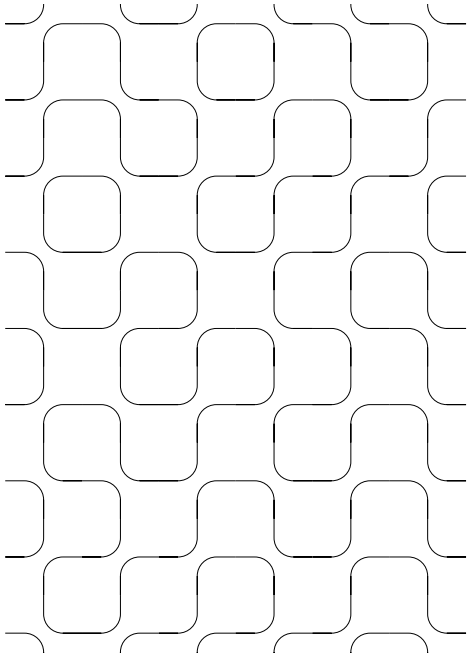}
\caption{Part of a typical configuration of the O$(1)$ loop model on a
$6\times \infty$ cylinder, periodic in the horizontal direction and
extending to infinity in both vertical directions.}\label{fig:O1}
\end{center}
\end{figure}

The determinant \eqref{dlth} has been evaluated exactly for the cases $\theta$ a multiple of $\dfrac{\pi}{3}$ \cite{lozenge,andrews,krattdet}. For general $\theta$, an approximate asymptotic expansion was obtained by Mitra and Nienhuis \cite{mtrloop}. They numerically studied the dense $O\haak{1}$ loop model \cite{loop} on a cylinder. The dense $O(1)$ loop model on square lattices can be defined as follows. At each vertex, the four edges meeting there are connected with equal probability in either of the two ways shown in Fig.\ \ref{fig:vrt}.
The set of edges connected to each other will then form closed loops with probability 1. In Fig.\ \ref{fig:O1} part of a typical example of a loop configuration is shown on a $6\times\infty$ lattice. Here we have imposed periodic boundary conditions in the horizontal direction which gives the lattice the topology of a cylinder with a circumference of $6$. If the circumference is chosen to be odd, then there will be "loop" spanning the entire length of the cylinder.

Mitra and Nienhuis conjectured an exact formula for the probability $P\haak{L,m}$ that a point on a cylinder with circumference $L$ is surrounded by $m$ loops \cite{mitradmtcs}. From that conjecture they were able to relate the generating function for the probability distribution to the determinant \eqref{dlth} \cite{mtrloop}. Defining
\begin{equation}\label{phidef}
\phi\haak{L,\theta}\equiv \sum_{m=0}^{L/2}P\haak{L,m}2^{m}\cos^{m}\haak{\theta}
\end{equation}
and
\begin{equation}\label{dltheta}
D\haak{L,\theta}\equiv\exp\haak{-\imath\theta L/2}\det_{1\leq r,s \leq 
L}\rhaak{\binom{r+s-2}{r-1}+\exp\haak{\imath\theta}\delta_{r,s}},
\end{equation}
the conjecture is
\begin{conjecture}\label{gen:con}
\begin{equation}\label{gen}
\begin{split}
\phi\haak{L,\theta}&= \frac{D\haak{L,\theta}}{A_{\text{HT}}\haak{L}^{2}}\text{ 
if } L\text{ is even,}\\
\phi\haak{L,\theta} &=\frac{1}{2\cos\haak{\frac{\theta}{2}}}\frac{D\haak{
L,\theta}}{A_{\text{HT}}\haak{
L}^{2}}\text{ if } L\text{ is odd.}
\end{split}
\end{equation}
\end{conjecture}
Here $A_{\text{HT}}\haak{L}$ is the number of $L\times L$ half-turn symmetric alternating sign matrices \cite{kup2,asmodd,rob}:
\begin{equation}\label{hl}
\begin{split}
A_{\text{HT}}\haak{L} &= 2\prod_{k = 1}^{\frac{L}{2}-1}\frac{3\left(k-1 \right)! k!\left(3 k-1 \right)!\left(3k+2 \right)!}
{4{\left(2k-1 \right)!}^2{\left(2k+1 \right) !}^2}\text{ if $L$ is even,}\\
A_{\text{HT}}\haak{L} &=\prod_{j = 1}^{\frac{L-1}{2}}\frac{4{j!}^2{\left(3j \right) !}^2} {3{\left( 2j \right) !}^4}\text{ if $L$ is odd.}
\end{split}
\end{equation}
We note that the function $D\haak{L,\theta}$ defined in Eq.\ \ref{dltheta} is an even function of $\theta$. This follows from a symmetry property of the coefficients of the characteristic polynomial of the Pascal matrix. If we put
\begin{equation}
\det_{1\leq r,s\leq L}\rhaak{\binom{r+s-2}{r-1}-x\delta_{r,s}}=\sum_{k=0}^{L}c_{k}x^{k},
\end{equation}
then $c_{L/2 + k} = \haak{-1}^{2k}c_{L/2 - k}$, see \cite{lunnon}. For reference later on in this paper, we note that the total number of $ n\times n $ alternating sign matrices, $A(n)$, is given by \cite{kup1,zeil}
\begin{equation}\label{al}
A\haak{n}=\prod_{k=0}^{n-1}\frac{\haak{3k+1}!}{\haak{k+n}!}.
\end{equation}

The asymptotic behavior of correlations in the O$(1)$ loop model can be calculated using
nonrigorous Coulomb gas techniques, see \cite{nijs,nienhuis,nienhuis2}. As shown in \cite{mtrloop}, this leads to the following conjecture.
\begin{conjecture}\label{exp:con}
\begin{equation}\label{exp}
\phi\haak{L,\theta}=\sum_{n=-\infty}^{\infty}C_{n}\haak{L,\theta}L^{-\dfrac{3\haaksp{\theta + 2n\pi}^{2}}{\haak{4\pi^{2}}}+\dfrac{1}{12}},
\end{equation}
where the functions $C_{n}\haak{L,\theta}$, which are not the same for odd and even $L$, have an asymptotic expansion in powers of $\dfrac{1}{L}$, such that the term proportional to $\dfrac{1}{L}$ vanishes.
\end{conjecture}
Conjectures \ref{gen:con} and \ref{exp:con} imply for $-\pi<\theta<\pi$:
\begin{equation}\label{dold} 
D\haak{L,\theta}=A\haak{\theta}\haak{\frac{3\sqrt{3}}{4}}^{L^{2}}L^{7/36-3\theta^{2}/\haak{4\pi^{2}}}\rhaak{1 + \mathcal{O}\haak{\frac{1}{L^{\min\rhaaksp{2,3\haaksp{1-\lhaak{\theta}/\pi}}}}}}.
\end{equation}
In Ref.\ \cite{mtrloop}, the amplitude $A\haak{\theta}$ was evaluated numerically. 
It was observed that:
\begin{conjecture}\label{oddiseven:con}
The amplitude $A\haak{\theta}$ in the expansion of $D\haak{L,\theta}$ is the same function for odd and even $L$.
\end{conjecture} 
Exact values for  $A\haak{\theta}$ were obtained for the cases $\theta$ a multiple of $\dfrac{\pi}{3}$ using the known exact evaluations of $D\haak{L,\theta}$.

The results of this paper were obtained by making two additional assumptions.
\begin{assumption}\label{anal:as}
The functions $C_{n}\haak{L,\theta}$ in \eqref{exp} are analytic in $\theta$. 
\end{assumption}
This was also mentioned in \cite{mtrloop}. This assumption, combined with the periodicity of $\phi\haak{L,\theta}$ implies that the $C_{n}\haak{L,\theta}$ are analytic continuations of each other
\begin{equation}
C_{n}\haak{L,\theta}=C_{0}\haak{L,\theta+2\pi n}.
\end{equation}
Another assumption we make is
\begin{assumption}\label{evenL:as}
The asymptotic expansion of $C_{0}\haak{L,\theta}$ in powers of $\dfrac{1}{L}$ contains only even powers of $\dfrac{1}{L}$.
\end{assumption}
This is supported by the known exact evaluations of $D\haak{L,\theta}$ for $\theta$ a multiple of $\dfrac{\pi}{3}$ and by numerical computations which we discuss in the next section.

Using Conjectures \ref{gen:con} and \ref{exp:con} and Assumptions \ref{anal:as} and \ref{evenL:as}, we were able to generate very accurate numerical data for the terms in the asymptotic expansion of $C_{0}\haak{L,\theta}$. This allowed us to guess the exact expressions in the asymptotic expansion up to relative order $L^{-14}$ to the leading term. The asymptotic expression for $D\haak{L,\theta}$ is given for both even and odd $L$ by
\begin{equation}
D\haak{L,\theta} = \sum_{n=-\infty}^{\infty}\haak{-1}^{n L}f\haak{L,\theta+2\pi n},
\end{equation}
where
\begin{equation}
\begin{split}
&f\haak{L,\theta}=\haak{\frac{2}{3}}^{\frac{1}{12}}\exp\haak{-5\zeta'\haak{-1}}\haak{\frac{3\sqrt{3}}{4}}^{L^{2}}L^{\frac{7}{36}-\frac{3\theta^{2}}{4\pi^{2}}}G\haak{1+\frac{\theta}{2\pi}}G\haak{1-\frac{\theta}{2\pi}}\\
&\times G\haak{\frac{4}{3}+\frac{\theta}{2\pi}}G\haak{\frac{4}{3}-\frac{\theta}{2\pi}}G\haak{\frac{2}{3}+\frac{\theta}{2\pi}}G\haak{\frac{2}{3}-\frac{\theta}{2\pi}}\exp\haak{\sum_{k=1}^{\infty}\frac{R_{2k}\haak{\theta}}{L^{2k}}}.
\end{split}
\end{equation}
Here $G\haak{z}$ is the Barnes G-function \cite{barnsasymp,barnesspec,barnes1, barnes2,barnes3}, and the $R_{2k}\haak{\theta}$ are polynomials of degree $2k+2$. For $k=1,\ldots,7$ they are given by Eq.\ \eqref{polr} in Section~\ref{sec:exact}. For odd $L$ this formula is implied by the validity of this formula for even $L$ and Conjecture~\ref{oddiseven:con}.

In the remainder of this paper we will discuss in detail how the above expression was obtained. In Section~\ref{sec:num} we explain how we obtained the numerical data for the terms in the asymptotic expansion. In Section~\ref{sec:exact} we discuss guessing exact expressions from the numerical data and how that led to our result. In Section~\ref{sec:loop} we apply the obtained asymptotic expansion to the loop model and compute the asymptotics for a number of correlations in this model.

\section{Generating high precision numerical data}\label{sec:num}
For convenience, we will first only consider the case of even $L$. We will return to the case of odd $L$ later on in this paper. From \eqref{gen}, \eqref{exp}, and \eqref{anal:as} it follows that
\begin{equation}\label{fit}
\frac{D\haak{L,\theta}}{A_{\text{HT}}^{2}\haak{L}}L^{\dfrac{3\theta^{2}}{\haak{4\pi^{2}}}-\dfrac{1}{12}}=\sum_{n=-\infty}^{\infty}\sum_{k=0}^{\infty}A_{k}\haak{\theta+2\pi n}L^{-3n^{2}-\dfrac{3n\theta}{\pi}-k},
\end{equation}
where the $A_{k}\haak{\theta}$ are even functions of $\theta$. To compute these functions, we generated the characteristic polynomial of the Pascal matrix up to $L=200$, which took a few days on an ordinary PC. Using these polynomials we could quickly evaluate the left-hand side of the above expression. This allowed us to extract the $A_{k}\haak{\theta}$ and their first few derivatives.

Let us first see what we can conclude from the exact expressions for $D\haak{L,\theta=n\frac{\pi}{3}}$, see Table~\ref{eval}.
\begin{table}
\begin{center}
\caption{Values for $D\haak{\theta}$ for $\theta$ a multiple of $\pi/3$. The expressions for $A\haak{L}$ and $A_{\text{HT}}\haak{L}$ are given in \eqref{al} and \eqref{hl}}\label{eval}
\begin{tabular}{|c|c|c|}\hline
$\theta$ & \multicolumn{2}{c|}{$D\haak{L,\theta}$}\\\hline
&even $L$&odd $L$\\\hline
0 & $\frac{A_{\text{HT}}\haak{2L}}{A\haak{L}}$ 
&$\frac{A_{\text{HT}}\haak{2L}}{A\haak{L}}$\\\hline
$\pi/3$ & $A_{\text{HT}}\haak{L}^{2}$ 
&$\sqrt{3}A_{\text{HT}}\haak{L}^{2}$\\\hline
$2 \pi/3$ & $A\haak{L}$ & $A\haak{L}$\\\hline
$\pi$ & $A\haak{\frac{L}{2}}^{4}$&0\\\hline
\end{tabular}
\end{center}
\end{table}
We can expand these expressions in powers of $\dfrac{1}{L}$ by rewriting the product of factorials in terms of Barnes G-functions \cite{barnsasymp,barnesspec,barnes1, barnes2,barnes3} and using the asymptotic expansion of the Barnes G-function, see e.g.\ Ref.\ \cite{mtrloop} for an efficient method to do this. We note that the asymptotic expansion of the Barnes G-function given in some papers contains a sign error, in particular in Refs.\ \cite{barnesspec,mtrloop}. The correct expansion is
\begin{equation}\label{bgas}
\begin{split}
\log\rhaak{G\haak{z+1}}=&z^{2}\haak{\half\log\haak{z}-\frac{3}{4}}+\half\log\haak{2\pi}z-
\frac{1}{12}\log\haak{z}+\zeta'\haak{-1}\\
\text{}&+\sum_{k=1}^{\infty}\frac{B_{2k+2}}{4k\haak{k+1}z^{2k}}.
\end{split}
\end{equation}
Here the $B_{k}$ are the Bernoulli numbers, and $\zeta'\haak{z}$ is the derivative of the zeta function. This can be derived using the methods presented in \cite{barnesspec}. A simple sign error leading to an incorrect minus sign in front of the summation over $k$ was made in that article. 

It turns out that the asymptotic expansions of the expressions in Table~\ref{eval} contain only even powers of $\dfrac{1}{L}$:
\begin{equation}\label{special}
\begin{split}
&\begin{split}
&\phi\haak{L,0}=\frac{3^{1/12}}{2^{5/36}}\exp\haak{-\zeta'\haak{-1}}\rhaakl{1+\frac{127}{5184 L^2}
 -\frac{2041055}{53747712 L^4}+\frac{107538127903}{835884417024 L^6}\vphantom{-\frac{272944577297197688875376083}{2794811034494209364066304 L^{12}}+\frac{26385460676926169502575757887765} {14488300402817981343319719936 L^{14}}}}\\ 
\text{}&-\frac{13294838545991999}{17332899271409664 L^8}+\frac{645434518069131955571} {89853749822987698176 L^{10}}\\
&\rhaakr{\vphantom{1+\frac{127}{5184 L^2}
 -\frac{2041055}{53747712 L^4}+\frac{107538127903}{835884417024 L^6}}\text{} -\frac{272944577297197688875376083}{2794811034494209364066304 L^{12}}+\frac{26385460676926169502575757887765} {14488300402817981343319719936 L^{14}}+\cdots}
\end{split}\\
&\phi\haak{L,\frac{\pi}{3}}=1\\
&\begin{split}
&\phi\haak{L,\frac{2\pi}{3}}=\frac{2^{31/36}}{3^{5/12}}\frac{\pi}{\Gamma\haak{1/3}^{2}}\exp\haak{-\zeta'\haak{-1}}\rhaakl{1+ \frac{7}{576 L^2} - \frac{23983}{663552 L^4} + \frac{16317695}{127401984 L^6}\vphantom{-\frac{19038476800109154745}{194775186325635072 L^{12}}+\frac{204450994938396835527815} {112190507323565801472 L^{14}}}}\\
\text{}&- \frac{225307455655}{293534171136 L^8} +\frac{1215802858094435}{169075682574336 L^{10}}\\
&\rhaakr{-\frac{19038476800109154745}{194775186325635072 L^{12}}+\frac{204450994938396835527815} {112190507323565801472 L^{14}}+\cdots\vphantom{1+ \frac{7}{576 L^2} - \frac{23983}{663552 L^4} + \frac{16317695}{127401984 L^6}}}
\end{split}\\
&\begin{split}
&\phi\haak{L,\pi}=\frac{2^{8/3}}{3}\frac{\pi^{2}}{\Gamma\haak{1/3}^{4}}\rhaakl{1- \frac{8}{81 L^2}  + \frac{464}{6561 L^4} - \frac{228352}{1594323 L^6}+ \frac{77553152}{129140163 L^8}\vphantom{- \frac{45379702784}{10460353203 L^{10}}+\frac{122234658136064}{2541865828329 L^{12}}-\frac{156017791843041280}{205891132094649 L^{14}}}}\\
&\rhaakr{\text{}- \frac{45379702784}{10460353203 L^{10}}+\frac{122234658136064}{2541865828329 L^{12}}-\frac{156017791843041280}{205891132094649 L^{14}}+\cdots\vphantom{1- \frac{8}{81 L^2}  + \frac{464}{6561 L^4} - \frac{228352}{1594323 L^6}+ \frac{77553152}{129140163 L^8}}}.
\end{split}
\end{split}
\end{equation}
The absence of odd powers of $\dfrac{1}{L}$ suggests that for general $\theta$,   $A_{k}\haak{\theta}=0$ for odd $k$. Our numerical results are consistent with this (the fact that $A_{1}\haak{\theta}=0$ is a Coulomb gas prediction). From \eqref{fit} we see that the $A_{2k}\haak{\theta+2\pi n}$ will also give rise to odd powers of $\dfrac{1}{L}$ when $\theta=0$ or $\theta=\dfrac{2\pi}{3}$ for odd $n$:
\begin{equation}\label{zer1}
\begin{split}
A_{2k}\rhaak{2\pi\haak{2n+1}}&=0,\\
A_{2k}\rhaak{\frac{2\pi}{3}+2\pi\haak{2n+1}}&=0.
\end{split}
\end{equation}
From the leading terms of the above expansions we can read-off $A_{0}\haak{\frac{\pi}{3}p}$ for $p=0,1,2,3$. Note that $A_{0}\haak{\pi}$ is half the leading term of $\phi\haak{L,\pi}$ because  $n=0$ and $n=-1$ contribute equally to the summation in \eqref{fit} when $\theta=\pi$. We can read-off some more $A_{k}\haak{\frac{\pi}{3}p}$ for $k>0$, but it is not possible to find more special cases of the form $A_{0}\haak{\frac{\pi}{3}p + 2\pi n}$ due to mixing with the $A_{k}\haak{\frac{\pi}{3}p}$. To calculate these values, we must lift the degeneracy between these terms, which can be done by considering the derivatives of \eqref{fit} w.r.t.\ $\theta$.

Our strategy to find $A_{0}\haak{\theta}$ was as follows:
\begin{enumerate}\label{program}
\item Using (derivatives of) Eq.\ \eqref{fit} obtain accurate fits for special values of $A_{0}\haak{\theta}$ and its derivatives.

\item Try to guess exact expressions for these approximate values.

\item Try to get information about the zeroes of $A_{0}\haak{\theta}$ from the fits.

\item \label{fndzer} Assuming that $A_{0}\haak{\theta}$ is of finite order, any guess about the location of all the zeroes (and their multiplicities) corresponds, by Hadamard's factorization theorem \cite{had}, to a guess for $A_{0}\haak{\theta}$ up to a factor $\exp\haak{\text{polynomial}}$. Such a guess can be easily checked by evaluating summations over the zeroes of the form $\sum_{n}\dfrac{m_{n}}{\haak{\alpha_{n} -\alpha}^{p}}$, where the $\alpha_{n}$ are the zeroes of $A_{0}\haak{\theta}$, the $m_{n}$ the multiplicities of the zeroes, and $p$ is an integer larger than the order of $A_{0}\haak{\theta}$. Such summations can be expressed in terms of the derivatives of $A_{0}\haak{\theta}$ at $\theta=\alpha$  as follows. The contour integral
\begin{equation}
\oint_{C\haak{R}}\frac{A_{0}'\haak{z}}{A_{0}\haak{z}}\frac{dz}{\haak{z-\alpha}^{p}},
\end{equation}
where $C\haak{R}$ is a circle with the origin as its center and radius $R$, will tend to zero for $R\longrightarrow\infty$ if $p$ is chosen larger than the order of $A\haak{\theta}$. It then follows from the residue theorem that
\begin{equation}\label{zerder}
\sum_{n}\frac{m_{n}}{\haak{\alpha_{n}-\alpha}^{p}} = -\frac{1}{\haak{p-1}!}\lhaakr{\frac{d^{p}}{dz^{p}}\log\rhaak{A_{0}\haak{z}}}_{z=\alpha}.
\end{equation}

If, using either numerical or exact expressions for the derivatives of $A_{0}\haak{\theta}$, we verify the validity of the above formula for some particular $\alpha$ and $p$, we can construct a candidate function for $A_{0}\haak{\theta}$ and verify that it fits the data. Note that it is theoretically possible for the last test to fail. E.g., it could be that $p$ is not larger than the order of $A_{0}\haak{\theta}$ and the guess about the zeroes is also wrong in such a way that \eqref{zerder} is satisfied by accident. 

\item \label{fail} If (\ref{fndzer}) fails, we could try to apply some transformation to $A_{0}\haak{\theta}$, e.g., by subtracting some function from it and consider the zeroes of the transformed function. Or we could abandon the attempt to find the function from its zeroes altogether and do some sort of brute force search instead.
\end{enumerate}

\section{Finding exact expressions from numerical data}\label{sec:exact}
An important result we obtained was the derivative of $A_{0}\haak{\theta}$ at $\theta=\dfrac{\pi}{3}$. The numerical value we obtained by fitting the derivative of \eqref{fit} was
\begin{equation}
A_{0}'\haak{\frac{\pi}{3}}\approx -0.31250232645180558377609866618987578223847934983479\ldots
\end{equation}
We conservatively estimated the accuracy of this number to be $\sim 30$ digits. To get an idea what the analytical expression of this number could possibly be like, let us consider the derivative of the function $\phi\haak{L,\theta}$. It follows from \eqref{exp} that the leading asymptotics of the derivative is
\begin{equation}
\phi'\haak{L,\frac{\pi}{3}}\sim A_{0}'\haak{\frac{\pi}{3}}-\frac{1}{2\pi}\log\haak{L}.
\end{equation}
Now, for finite $L$ the value $\phi'\haak{L,\frac{\pi}{3}}$ is just an algebraic number. The fact that on the right-hand side there is a logarithmic term in $L$ strongly suggests that the "constant term"  $A_{0}'\haak{\frac{\pi}{3}}$ contains a term proportional to Euler's constant with the same amplitude as the coefficient in front of the $\log\haak{L}$. If we subtract this contribution of Euler's  constant from our fit to $A_{0}'\haak{\frac{\pi}{3}}$ we get
\begin{equation}
A_{0}'\haak{\frac{\pi}{3}}+\frac{\gamma}{2\pi}\approx 
-0.22063560015265159339645643211799769082485006927415\ldots
\end{equation}
It is not so clear what this number is, but perhaps we should multiply it by $\pi$ because the term $\log\haak{L}+\gamma$ appears with a factor $\dfrac{1}{\haak{2\pi}}$:
\begin{equation}
\begin{split}
\pi\rhaak{A'\haak{\frac{\pi}{3}}+\frac{\gamma}{2\pi}}\approx &
-0.69314718055994530941723212145817656757963215693787\ldots\\
\text{This looks familiar:}\\
-\log\haak{2}=&-0.69314718055994530941723212145817656807550013436026\ldots
\end{split}
\end{equation}
So, we arrive at the following conjecture:
\begin{equation}
A_{0}'\haak{\frac{\pi}{3}}=
-\frac{\gamma}{2\pi } - \frac{\log\haak{2}}{\pi}.
\end{equation}

For general $\theta$:
\begin{equation}
\phi'\haak{L,\theta}\sim A_{0}'\haak{\theta}-\frac{3\theta}{2\pi^{2}}A_{0}\haak{\theta}\log\haak{L}.
\end{equation}
And we thus expect that $\dfrac{A_{0}'\haak{\theta}}{A_{0}\haak{\theta}}$ contains a term $-\dfrac{3\theta}{\haak{2\pi^{2}}}\gamma$. After subtracting this contribution we may not be able to recognize the remaining term as easily as in the case of $\theta=\dfrac{\pi}{3}$. The problem we need to solve is in general: We suspect that some number $x$ which we know to limited precision is a rational linear combination of some constants $y_{1}\ldots y_{n}$. We want to know if this is true, in which case we want to find the linear combination. This type of problem can be solved using so-called integer relation algorithms. We used the Lenstra-Lenstra-Lov\'{a}sz lattice reduction algorithm (LLL algorithm) \cite{lll} for this purpose, implemented in Mathematica by the function LatticeReduce. In Appendix~\ref{apl3} we give a brief description of this algorithm and explain how it can be used as an integer relation algorithm.

Using the LLL algorithm were able to find more exact values for the logarithmic derivative of $A_{0}\haak{\theta}$, such as
\begin{equation}
\begin{split}
\frac{A_{0}'\haak{\frac{2\pi}{3}}}{A_{0}\haak{\frac{2\pi}{3}}}&=-\frac{-2\gamma+3\log\haak{3}}{2\pi},\\
\frac{A_{0}'\haak{\pi}}{A_{0}\haak{\pi}}&=-\frac{-3\gamma-6\log\haak{2}+3\log\haak{3}}{2\pi}.\\
\end{split}
\end{equation}
The pattern of zeroes was not very clear at this stage. Fitting to the second derivative of \eqref{fit} at $\theta=0$ yielded very small coefficients for the $\log\haak{L}^{2}L^{-p}$ terms,  consistent with these terms being exactly zero. This suggests that there are zeroes at $2\pi n$ for $n\neq 0$. Fitting to the first and second derivative of \eqref{fit} at $\theta=\dfrac{2\pi}{3}$ yielded very small coefficients for the $\log\haak{L} L^{-p}$ and $\log\haak{L}^{2}L^{-p}$ terms, respectively. This then suggests additional zeroes at $2\pi n\pm \dfrac{2\pi}{3}$. Note that for odd $n$ this coincides with \eqref{zer1}.

As we can see from \eqref{special}, the $A_{k}\haak{\theta}$ rapidly diverge as a function of $k$. This makes it difficult to extract accurate values for $A_{0}\haak{\theta}$ for $\theta\gtrsim 3\pi$. We could only verify that the derivatives at $2\pi$, $\dfrac{4\pi}{3}$ and $\dfrac{8\pi}{3}$ were nonzero. Clearly it is not possible for all the zeroes to be simple zeroes because then $A_{0}\haak{\theta}$ would have to be a trigonometric function up to a polynomial factor, which is inconsistent with the results we found so far. E.g.\ $A_{0}\haak{\theta}$ cannot be a product of functions like $\Gamma\haak{a+b\theta}\Gamma\haak{a-b\theta}$. A product of functions of the form $\Gamma\haak{a+b\theta^{2}}$ is consistent with the derivatives we found, but is not consistent with the equal spacing of the zeroes. A product of Barnes G-functions is a more plausible possibility, however, then the multiplicities of the zeroes must increase linearly, but that is not what we see at the first few zeroes. 

We proceeded with attempting to find expressions for higher derivatives. The presence of Euler's constant in the first logarithmic derivative suggests that in higher (logarithmic) derivatives we can expect to find polygamma functions. Using the LLL algorithm we found
\begin{equation}
\frac{A''_{0}\haak{0}}{A_{0}\haak{0}}=-\frac{3}{2{\pi }^2}-\frac{3\gamma}{2{\pi }^2} -\frac{3\log\haak{3}}{2{\pi }^2}+\frac{\psi_{1}\haak{\frac{1}{3}}}{6{\pi}^2}- \frac{\psi_{1}\haak{\frac{2}{3}}}{6{\pi}^2},
\end{equation}
where $\psi_{p}\haak{z}$ is the polygamma function of order $p$.
If the order of $A_{0}\haak{\theta}$ is $2$ or higher then we need to consider higher derivatives to compute summations over the zeroes. If the order of $A_{0}\haak{\theta}$ is less than 4, then $\sum_{n}\dfrac{m_{n}}{\alpha_{n}^{4}}$ is given by $-4$ times the coefficient of $\theta^{4}$ in the series expansion of $\log\rhaak{A_{0}\haak{\theta}}$ around $\theta=0$.
 
After some trial and error we found using the LLL algorithm the expression:
\begin{equation}\label{sm4}
\frac{1}{4!}\lhaakr{\frac{d^{4}}{d\theta^{4}}\log\rhaak{A_{0}\haak{\theta}}}_{\theta=0} = \frac{1}{576\pi^{4}}\rhaak{\psi_{3}\haak{\frac{1}{3}}-\psi_{3}\haak{\frac{2}{3}}}-\frac{27}{32\pi^{4}}\zeta\haak{3}.
\end{equation}
We can reproduce this result if we assume that all the zeroes of $A_{0}\haak{\theta}$ are at $2\pi n$ and $2\pi n\pm 2\pi/3$ and have a multiplicity of $\lhaak{n}$.
The first three positive zeroes at $\dfrac{4\pi}{3}$, $2\pi$ and $\dfrac{8\pi}{3}$ then have a multiplicity of 1 as we observed earlier. To see that this works, let us write the summation over the zeroes as
\begin{equation}
\begin{split}
\sum_{n}\frac{m_{n}}{\alpha_{n}^{4}}&=\frac{1}{8\pi^{4}}\rhaak{\sum_{n=1}^{\infty}\frac{1}{n^{3}} + \sum_{n=1}^{\infty}\frac{n}{\haak{n-\frac{1}{3}}^{4}}+\sum_{n=0}^{\infty}\frac{n}{\haak{n+\frac{1}{3}}^{4}}}\\
&=\frac{1}{8\pi^{4}}\rhaak{\sum_{n=1}^{\infty}\frac{1}{n^{3}} + \sum_{n=1}^{\infty}\frac{1}{\haak{n-\frac{1}{3}}^{3}}+\sum_{n=0}^{\infty}\frac{1}{\haak{n+\frac{1}{3}}^{3}}}\\
\text{}&+\frac{1}{24\pi^{4}}\rhaak{\sum_{n=1}^{\infty}\frac{1}{\haak{n-\frac{1}{3}}^{4}}-\sum_{n=0}^{\infty}\frac{1}{\haak{n+\frac{1}{3}}^{4}}}.
\end{split}
\end{equation}
We can simplify this expression further using
\begin{equation}
\sum_{n=1}^{\infty}\frac{1}{n^{3}} + \sum_{n=1}^{\infty}\frac{1}{\haak{n-\frac{1}{3}}^{3}}+\sum_{n=0}^{\infty}\frac{1}{\haak{n+\frac{1}{3}}^{3}}=\sum_{n=1}^{\infty}\frac{1}{\haak{\frac{n}{3}}^{3}}=27\zeta\haak{3},
\end{equation}
and the identity:
\begin{equation}
\sum_{n=0}^{\infty}\frac{1}{\haak{n+z}^{p}}=\frac{\haak{-1}^{p}}{\haak{p-1}!}\psi_{p-1}\haak{z},
\end{equation}
which allows us to write
\begin{equation}
\sum_{n=1}^{\infty}\frac{1}{\haak{n-\frac{1}{3}}^{4}}=\sum_{n=0}^{\infty}\frac{1}{\haak{n+\frac{2}{3}}^{4}}=\frac{1}{6}\psi_{3}\haak{\frac{2}{3}},
\end{equation}
and
\begin{equation}
\sum_{n=0}^{\infty}\frac{1}{\haak{n+\frac{1}{3}}^{4}}=\frac{1}{6}\psi_{3}\haak{\frac{1}{3}}.
\end{equation}
These simplifications yield the expression:
\begin{equation}
\sum_{n}\frac{m_{n}}{\alpha_{n}^{4}}=\frac{27}{8\pi^{4}}\zeta\haak{3}+\frac{1}{144\pi^{4}}\rhaak{\psi_{3}\haak{\frac{2}{3}} -\psi_{3}\haak{\frac{1}{3}}}.
\end{equation}
We see from \eqref{sm4} that this is indeed the same as $-4$ times the coefficient of $\theta^{4}$ of $\log\rhaak{A_{0}\haak{\theta}}$.

The function $G\haak{1+z}$ has zeroes at the negative integers $z=-n$ with multiplicity $n$ and is of second order. The conjecture about the zeroes of $A_{0}\haak{\theta}$ and the assumption that the order of $A_{0}\haak{\theta}$ is less than four is thus equivalent to the statement: There exist constants $a$ and $b$ such that
\begin{equation}
\begin{split}
A_{0}\haak{\theta}=&\exp\haak{a+b\theta^{2}}G\haak{1+\frac{\theta}{2\pi}}G\haak{1-\frac{\theta}{2\pi}}G\haak{\frac{4}{3}+\frac{\theta}{2\pi}}G\haak{\frac{4}{3}-\frac{\theta}{2\pi}}\\
&\times G\haak{\frac{2}{3}+\frac{\theta}{2\pi}}G\haak{\frac{2}{3}-\frac{\theta}{2\pi}}.
\end{split}
\end{equation}
We verified that this fits the data exactly. We found
\begin{equation}
\begin{split}
A_{0}\haak{\theta}=&2^{11/36}3^{1/36}{\pi}^{1/3}\exp\haak{-\frac{19}{3}\zeta'\haak{-1}}{\Gamma\haak{\frac{1}{6}}}^{-\frac{2}{3}}G\haak{1+\frac{\theta}{2\pi}}G\haak{1 -\frac{\theta}{2\pi}}\\
\text{}\times&G\haak{\frac{2}{3}+\frac{\theta}{2\pi}}G\haak{\frac{2}{3}-\frac{\theta}{2\pi}}G\haak{\frac{4}{3}+\frac{\theta}{2\pi}}G\haak{\frac{4}{3} - \frac{\theta}{2\pi}}.
\end{split}
\end{equation}

Next, we turned our attention to the $A_{2k}\haak{\theta}$. We found that the ratios $\dfrac{A_{2k}\haak{\theta}}{A_{0}\haak{\theta}}$ appeared to be polynomials of degree $4k$. We observed that the summation over $k$ in \eqref{fit} can be simplified as follows. If we define the polynomials $Q_{2k}\haak{\theta}$ as
\begin{equation}
A_{0}\haak{\theta}\exp\rhaak{\sum_{k=0}^{\infty}\frac{Q_{2k}\haak{\theta}}{L^{2k}}}\equiv\sum_{k=0}^{\infty}\frac{A_{2k}\haak{\theta}}{L^{2k}}, 
\end{equation}
then $Q_{2k}\haak{\theta}$ is of degree $2k+2$. Our results so far can be summarized as follows. For even $L$, the function $\phi\haak{L,\theta}$ is given by an asymptotic expansion of the form:
\begin{equation}\label{evrslt}
\phi\haak{L,\theta}=\sum_{n=-\infty}^{\infty}A_{0}\haak{L,\theta+2\pi n}L^{-\dfrac{3\haaksp{\theta + 2n\pi}^{2}}{\haak{4\pi^{2}}}+\dfrac{1}{12}}\exp\rhaak{\sum_{k=0}^{\infty}\frac{Q_{2k}\haak{\theta+2\pi n}}{L^{2k}}}.
\end{equation}
What about odd $L$? In Ref.\ \cite{mtrloop} it was observed that the coefficient of the  $L^{-\dfrac{3\theta^{2}}{\haak{4\pi^{2}}}+\dfrac{1}{12}}$ term in $\phi\haak{L,\theta}$ for odd $L$ is $A_{0}^{\text{odd}}\haak{\theta}=\frac{\sqrt{3}}{2\cos\haaksp{\dfrac{\theta}{2}}}A_{0}\haak{\theta}$. An equation of the same form as \eqref{evrslt} for odd $L$ would thus imply that there exist polynomials $Q^{\text{odd}}_{2k}\haak{\theta}$ such that
\begin{equation}\label{odrslt}
\begin{split}
\phi\haak{L,\theta}=\frac{\sqrt{3}}{2\cos\haak{\frac{\theta}{2}}}\sum_{n=-\infty}^{\infty}&\haak{-1}^{n} A_{0}\haak{L,\theta+2\pi n}L^{-\dfrac{3\haaksp{\theta + 2n\pi}^{2}}{\haak{4\pi^{2}}}+\dfrac{1}{12}}\\
&\times\exp\rhaak{\sum_{k=0}^{\infty}\frac{Q^{\text{odd}}_{2k}\haak{\theta+2\pi n}}{L^{2k}}}
\end{split}
\end{equation}
for odd $L$. From \eqref{gen} we see that for even $L$, the function $D\haak{L,\theta}$ is obtained by multiplying $\phi\haak{L,\theta}$ by $A_{\text{HT}}\haak{L}^{2}$, while for odd $L$ we must multiply by $2\cos\haak{\frac{\theta}{2}}A_{\text{HT}}\haak{L}^{2}$. As was observed in \cite{mtrloop}, this leads to the same result for the leading term of $D\haak{L,\theta}$  because $A_{\text{HT}}\haak{L}$ for odd $L$ is $3^{-1/4}$ times $A_{\text{HT}}\haak{L}$ for even $L$ in the large $L$ limit. Our results therefore imply that the asymptotics for $D\haak{L,\theta}$ for both even and odd $L$ are given by the formula:
\begin{equation}\label{sum}
D\haak{L,\theta} = \sum_{n=-\infty}^{\infty}\haak{-1}^{n L}f\haak{L,\theta+2\pi n}
\end{equation}
where
\begin{equation}\label{result}
\begin{split}
&f\haak{L,\theta}=\haak{\frac{2}{3}}^{\frac{1}{12}}\exp\haak{-5\zeta'\haak{-1}}\haak{\frac{3\sqrt{3}}{4}}^{L^{2}}L^{\frac{7}{36}-\frac{3\theta^{2}}{4\pi^{2}}}G\haak{1+\frac{\theta}{2\pi}}G\haak{1-\frac{\theta}{2\pi}}\\
&\times G\haak{\frac{4}{3}+\frac{\theta}{2\pi}}G\haak{\frac{4}{3}-\frac{\theta}{2\pi}}G\haak{\frac{2}{3}+\frac{\theta}{2\pi}}G\haak{\frac{2}{3}-\frac{\theta}{2\pi}}\exp\haak{\sum_{k=1}^{\infty}\frac{R_{2k}\haak{\theta}}{L^{2k}}}.
\end{split}
\end{equation}
The polynomials $R_{2k}\haak{\theta}$ for $k=1,\ldots, 7$ are given by
\begin{equation}\label{polr}
\begin{split}
R_{2}\haak{\theta} &= \frac{77}{15552} + \frac{7 {\theta }^2}{144 {\pi }^2} - \frac{11 {\theta }^4}{64 {\pi }^4}\\
R_{4}\haak{\theta} &= -\frac{245}{559872} - 
\frac{157 {\theta }^2}{12960 {\pi }^2} - \frac{29 {\theta }^4}{1152 {\pi }^4} + \frac{181 {\theta }^6}{1280 {\pi }^6}\\
R_{6}\haak{\theta} &= \frac{1103}{40310784} - \frac{1349
{\theta }^2}{244944 {\pi }^2} + \frac{3599 {\theta }^4}{31104 {\pi }^4} - \frac{989 {\theta }^6}{6912 {\pi }^6} - \frac{3275 {\theta }^8}{14336 {\pi }^8}\\
R_{8}\haak{\theta} &= \frac{793135}{4353564672} +
\frac{116807 {\theta }^2}{2099520 {\pi }^2} - \frac{101009 {\theta }^4}{279936 {\pi }^4} - \frac{47479 {\theta }^6}{622080 {\pi }^6} + \frac{43171 {\theta }^8}{36864 {\pi }^8} + \frac{61621 {\theta }^{10}}{122880 {\pi }^{10}}\\
 & \hspace{-29 pt}\begin{split}
 R_{10}\haak{\theta} = &-\frac{93651593}{130606940160} - \frac{3740009 {\theta }^2}{10392624 {\pi }^2} + \frac{1868083 {\theta }^4}{1399680 {\pi }^4} + \frac{301091 {\theta }^6}{93312 {\pi }^6} - \frac{1858513 {\theta }^8}{276480 {\pi }^8}\\
&\text{} - \frac{1239773 {\theta }^{10}}{184320 {\pi }^{10}} - \frac{1184171 {\theta }^{12}}{901120 {\pi }^{12}}
\end{split}\\
& \hspace{-29 pt}\begin{split}
R_{12}\haak{\theta} =&\frac{2884889645}{940369969152} + \frac{68061091601 {\theta }^2}{23213342880 {\pi }^2} - \frac{110018569 {\theta }^4}{22674816 {\pi }^4} - \frac{754814143 {\theta }^6}{16796160 {\pi }^6}\\
&\text{}+ \frac{454871621 {\theta }^8}{10450944 {\pi }^8} + \frac{931652293 {\theta }^{10}}{9953280 {\pi }^{10}} + \frac{31193731 {\theta }^{12}}{884736 {\pi }^{12}} + \frac{23057581 {\theta }^{14}}{5963776 {\pi }^{14}}
\end{split}\\
& \hspace{-29 pt}\begin{split}
R_{14}\haak{\theta} = &-\frac{2213492219141}{135413275557888} - \frac{2471502605 {\theta }^2}{76527504 {\pi }^2} - \frac{83019415531 {\theta }^4}{7142567040 {\pi }^4} + \frac{30869634919 {\theta }^6}{45349632\, {\pi }^6}\\
& \text{}- \frac{10100916773 {\theta }^8}{44789760 {\pi }^8} - \frac{96936237491 {\theta }^{10}}{62705664 {\pi }^{10}} - \frac{35619671389 {\theta }^{12}}{39813120 {\pi }^{12}} - \frac{105293315 {\theta }^{14}}{589824{\pi }^{14}}\\
 & \text{}- \frac{453005291 {\theta }^{16}}{36700160 {\pi }^{16}}
\end{split}
\end{split}
\end{equation}

We note that the summation in the exponent in \eqref{result} is a divergent asymptotic expansion of some unknown function. If we truncate this expansion at some finite order, then the summation over $n$ in \eqref{sum} will diverge. However, since the summation over $n$ converges very fast (independently of $L$), this is not an issue when doing numerical computations.

\section{Application to the O$(1)$ loop model and bond percolation}\label{sec:loop}
\begin{figure}[ht]
\begin{center}
\includegraphics[width=0.5\textwidth]{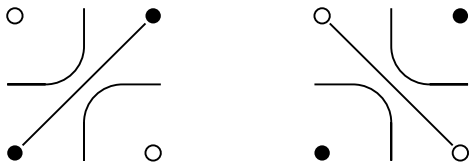}
\end{center}
\caption{The mapping of a loop configuration to a bond configuration of the corresponding bond percolation problem. A bond is either put on an edge of the square lattice formed by the 
$\bullet$ or on the dual edge 
orthogonal to it on the square lattice formed by the $\circ$.}\label{fig:perc}
\end{figure}

In this section, we consider the leading asymptotics of some correlations in the O$\haak{1}$ loop model and bond percolation model on the cylinder that can now be evaluated exactly. The bond percolation model at the critical point is related to the dense O$(1)$ loop model via the bijection shown in 
Fig.\ \ref{fig:perc}. 
\setlength{\unitlength}{\textwidth}
We can see from this figure that the loops of the dense O$(1)$ loop model correspond to boundaries of percolation clusters. From \eqref{phidef} we can see that the generating function $\phi\haak{L,\theta=\dfrac{\pi}{2}}$ equals the probability that a point in the loop model is not surrounded by any loops. This probability thus corresponds to the probability in the percolation model that a point is on a cluster that wraps around the cylinder for even $L$, while for odd $L$ it gives the probability that the point is on a cluster that spans the cylinder. This probability was given in \cite{mitradmtcs} for even $L$ as $P\approx 0.81099753 L^{-5/48}\rhaak{1+\mathcal{O}\haak{L^{-3/2}}}$. Using \eqref{evrslt} we can evaluate the leading asymptotics as
\begin{equation}
P=\frac{2^{23/72}}{3^{5/48}}\frac{\pi^{1/4}\exp\rhaak{-1/4\zeta'\haak{-1}}}{\sqrt{\Gamma\haak{1/4}}}L^{-5/48}\rhaak{1+\mathcal{O}\haak{L^{-3/2}}}.
\end{equation}
It follows from \eqref{evrslt} that the leading asymptotics for the probability that a point is on a cluster that spans the cylinder is $\sqrt{3/2}$ times the above expression.

Next, we consider the number of loops surrounding a point. By repeatedly differentiating both sides of Eq.\ \eqref{phidef} at $\theta=\dfrac{\pi}{3}$, we can express the moments of $P\haak{L,m}$ in terms of derivatives of $\phi\haak{L,\theta}$ at $\theta=\dfrac{\pi}{3}$. These derivatives can be evaluated using the derivatives of \eqref{evrslt} and \eqref{odrslt} at $\theta=\dfrac{\pi}{3}$. Denoting the number of loops surrounding a point by $N$, we have
\begin{equation}
\begin{split}
\gem{N}&=\frac{1}{2\sqrt{3}\pi}\rhaak{\gamma+\log\haak{4L}}\rhaak{1+\mathcal{O}\haak{L^{-2}}}\text{ for even } L,\\
\gem{N}&=\ahaak{-\frac{1}{6}+\frac{1}{2\sqrt{3}\pi}\rhaak{\gamma+\log\haak{4L}}}\rhaak{1+\mathcal{O}\haak{L^{-2}}}\text{ for odd } L.
\end{split}
\end{equation}
Using the first and second derivatives at $\theta=\dfrac{\pi}{3}$ the fluctuation in $N$ can be calculated. This is given by
\begin{equation}
\begin{split}
\gem{N^{2}}-\gem{N}^{2}&=\ahaak{-\frac{1}{9}-\frac{1+\log\haak{3}}{2\pi^{2}}+\frac{\psi_{1}\haak{1/6}}{18\pi^{2}} + \rhaak{\frac{2}{3\sqrt{3}\pi}-\frac{1}{2\pi^{2}}}\rhaak{\gamma+\log\haak{4L}}}\\
&\times\rhaak{1+\mathcal{O}\haak{L^{-2}}}\text{ for even } L, \\
\gem{N^{2}}-\gem{N}^{2}&=\ahaak{-\frac{2}{9}-\frac{1+\log\haak{3}}{2\pi^{2}}+\frac{\psi_{1}\haak{1/6}}{18\pi^{2}} + \rhaak{\frac{2}{3\sqrt{3}\pi}-\frac{1}{2\pi^{2}}}\rhaak{\gamma+\log\haak{4L}}}\\
&\times\rhaak{1+\mathcal{O}\haak{L^{-2}}}\text{ for odd } L. 
\end{split}
\end{equation}
\section{Discussion}
We have obtained the exact asymptotics of the characteristic polynomial of the Pascal matrix. The form of the asymptotics was partially known due to a conjecture relating $D\haak{L,\theta}$ to $\phi\haak{L,\theta}$. This was used to obtain accurate numerical data for the amplitude of the leading power of $L$. The LLL lattice reduction algorithm was used to guess exact expressions for special values of the amplitude which ultimately enabled us to find exact expressions for the amplitude for all $\theta$ and the amplitudes of the subleading terms up to relative order $L^{-14}$ to the leading term. 

We have applied this result to the O$(1)$ loop model and the related bond percolation model on the cylinder. Asymptotic expressions for the average of the number of loops surrounding a point and the fluctuation in this number were obtained. We also obtained the leading asymptotics of the probability that a point is not surrounded by any loops, which for even $L$ corresponds to the probability that a point is on a cluster that wraps around the cylinder, while for odd $L$ it gives the probability that a point is on a cluster that spans the length of the cylinder.

The methods presented here can likely be used to obtain exact asymptotic expressions for a large class of binomial determinants. E.g., in \cite{lozenge}, exact expressions for the determinant $\det_{1\leq r,s\leq L}\rhaak{\binom{r+s+m-2}{r-1}+\exp\haak{i\theta}\delta_{r,s}}$ for arbitrary integer $m$ and $\theta$ a multiple of $\pi/3$ were obtained. The asymptotics for general $\theta$ appears to have a similar structure as for the case $m=0$ investigated in this article.

\section{Acknowledgments}
I thank Bernard Nienhuis for useful conversations and the anonymous referees for their comments which helped to improve this paper.

\appendix

\section{Solving integer relation problems using the LLL algorithm}\label{apl3}
The LLL algorithm solves the following problem. Suppose we are given $n$ linear independent vectors $\bfm{v}_{1}\ldots\bfm{v}_{n}$ in $\mathbb{R}^{d}$ with $d\geq n$. Consider the set of all integer linear combinations of the $\bfm{v}_{j}$. Such a set is called a lattice and the $\bfm{v}_{j}$ are a basis of the lattice. The basis of a lattice is not unique in general. Given the $\bfm{v}_{j}$, the LLL algorithm finds short, approximately orthogonal, basis vectors for the lattice spanned by the $\bfm{v}_{j}$.

In an integer relation problem, one is given a real number $x$ which one wishes to express in terms of real constants $y_{k}$:
\begin{equation}
x = \sum_{k}^{n}r_{k}y_{k},
\end{equation}
where the $r_{k}$ are rational numbers. In our case, $x$ is known to limited precision while the $y_{k}$ are known to arbitrary precision. We want to know if for given $x$ and $y_{k}$ such a relation exists, in which case we want to find the $r_{k}$. The LLL algorithm can be used to solve this problem as follows. We first multiply our number $x$ and the constants $y_{1}\ldots y_{n}$ by some power of ten and take the floor to get integers:
\begin{equation}
\begin{split}
y'_{i}&\equiv \floor{10^{p}y_{i}},\\
x'&\equiv \floor{10^{p}x}.
\end{split}
\end{equation}
We choose $p$ such that $x'$ contains only those digits of $x$ that we know to be correct. We then define the basis vectors:
\begin{equation}
\begin{split}
\bfm{v}_{i}&\equiv\bfm{e}_{i}+y'_{i}\bfm{e}_{n+2}\text{ for } 1\leq i\leq n,\\
\bfm{v}_{n+1}&\equiv\bfm{e}_{n+1}+x'\bfm{e}_{n+2}.
\end{split}
\end{equation}
where $\bfm{e}_{r}$ is the $r$th unit vector of $\mathbb{R}^{n+2}$. We take these $n+1$ vectors $\bfm{v}_{1}\ldots \bfm{v}_{n+1}$ as the input of the LLL algorithm. The output will be a set of short basis vectors for the lattice spanned by the $\bfm{v}_{i}$. If there is a short basis vector $\bfm{b}$ in the output for which the $\haak{n+2}$nd component is $\mathcal{O}\haak{1}$ in $n$ and the $\haak{n+1}$st component is nonzero, then we have very likely found an exact relation between the $y_{i}$ and $x$:
\begin{equation}
\sum_{k=1}^{n}b_{k}y_{k} + b_{n+1}x = 0.
\end{equation}
The larger the number $n$ of constants is, the larger we need to choose $p$ to be able to detect a relation.

The LLL algorithm is not the best available integer relation algorithm. The PSLQ algorithm \cite{pslq} is faster and will yield lower bounds on the sizes of the coefficients if no relation is found. However, unlike the PSLQ algorithm, the LLL algorithm can also be used to find an unknown function given the action of a finite number of linear functionals. This would be useful in the event we need to move on to point~\ref{fail} of the program described on p.\ \pageref{program}. Suppose an unknown function $f:S_{1}\rightarrow S_{2}$ for arbitrary vector spaces $S_{1}$ and $S_{2}$ satisfies the relations:
\begin{equation}
L_{r}f=x_{r}
\end{equation}
for $1\leq r\leq m$ where the $L_{r}$ are linear functionals and the $x_{r}$ are real numbers. If $f$ is suspected to be a rational linear combination of functions $g_{k}$:
\begin{equation}
f = \sum_{k=1}^{n}r_{k}g_{k},
\end{equation}
where $n>m$, we can use the LLL algorithm to find the $r_{k}$ if they exist. We define
\begin{equation}
\begin{split}
y_{k,r}&\equiv L_{r}g_{k},\\
y'_{k,r}&= \floor{10^{p}y_{k,r}},\\
x'_{r}&= \floor{10^{p}x_{r}}
\end{split}
\end{equation}
for $1\leq k\leq n$ and $1\leq r\leq m$, where $p$ is limited by the accuracy of the $x_{r}$.
We then take as input for the LLL algorithm the basis vectors defined as
\begin{equation}
\begin{split}
\bfm{v}_{i}&\equiv\bfm{e}_{i}+\sum_{r=1}^{m}y'_{i,r}\bfm{e}_{n+1+r}\text{ for } 1\leq i\leq n,\\
\bfm{v}_{n+1}&\equiv\bfm{e}_{n+1}+\sum_{r=1}^{m}x_{r}'\bfm{e}_{n+1+r}.
\end{split}
\end{equation}
If a short output basis vector $\bfm{b}$ has a nonzero $\haak{n+1}$st component and the components $\haak{n+1+r}$ for $r=1,\ldots, m$ are all $\mathcal{O}\haak{1}$ in $n$, then that vector very likely defines an exact relation between $f$ and the $g_{k}$:
\begin{equation}
\sum_{k=1}^{n}b_{k}g_{k} + b_{n+1}f = 0,
\end{equation}
provided $p$ was chosen large enough.

\end{document}